\documentclass[12pt]{iopart}
\usepackage{color}
\usepackage[colorlinks=true,linkcolor=red]{hyperref}
\usepackage{iopams}
\begin{document}
\title{Holographic principle in spacetimes with extra spatial dimensions}
\author{P. Midodashvili}
\address{Tskhinvali State University, 2 Besiki Str., Gori 1400,
Georgia} \ead{pmidodashvili@yahoo.com}
\begin{abstract}
F. Scardigli and R. Casadio have considered uncertainty principles
which take into account the role of gravity and possible existence
of extra spatial dimensions. They have argued that the predicted
number of degrees of freedom enclosed in a given spatial volume
matches the holographic counting only for one of the available
generalization and without extra dimensions. Taking into account the
additional inevitable source of uncertainty in distance measurement,
which is missed in their approach, we show that the holographic
properties of the proposed uncertainty principle is recovered in the
models with extra spatial dimensions.
\end{abstract}
\maketitle

As it is well known the holographic principle is claimed to apply to
all gravitational systems. Using the error due to gravitational
curvature, originating from the Schwarzschild metric surrounding the
clock, the authors of article \cite{Scar-Cas} have investigated the
holographic properties of the generalized uncertainty principles
extended to the brane world scenarios. They argued that when extra
spatial dimensions are admitted, the holography is destroyed. In
this article we investigate this issue in more detail. From our
point of view the method of distance uncertainty estimation proposed
in \cite{Scar-Cas} misses one additional inevitable source of
uncertainty in distance measurement, and because of this fact the
method accidentally works in the case of $4$-dimensional spacetime
and gives an incorrect result in the case of extra dimensions.

Let us proceed to our arguments. In article \cite{Scar-Cas} the
authors have considered the measuring of a distance $l$ in the ADD
model \cite{ADD-Model} in the following four different cases: $1)~0
< L < r_{S\left( {4+n} \right)}  < a < l~, ~~2)~0 < r_{S\left( {4 +
n} \right)}  < L < a < l~, ~~3 )~0 < r_{S\left( {4 + n} \right)}  <
a < L < l~,~~4)~0 < r_{S\left( {4 + n} \right)} < a < l < L~,$ where
$L$ is the size of the compact extra dimensions in the ADD-model,
$a$ is the size of the spherically symmetric clock with mass $m$ and
$r_{S\left( {4+n} \right)}$ is the clock's Schwarzschild radius. The
authors of \cite{Scar-Cas} argued that in cases 3 and 4 the
holographic principle is destroyed for $n \ge 1$. In what follows
let us consider in more detail case $4$ (case $3$ can be considered
in a similar manner). In this case, when characteristic lengths
involved in the measurement (i.e., clock size $a$ and distance $l$
to be measured) are smaller than the radius $L$ of the compact
dimensions, the measurement process should be insensitive to the
brane and the boundary conditions in the $n$ transverse dimensions,
and so the measurement can be considered as a measuring process
conducted in a purely $(4+n)$-dimensional spacetime. In
\cite{Scar-Cas} the total uncertainty in distance measurement was
presented in the form: $\delta l_{tot}=\delta l_{QM}+\delta l_{C}$,
where $\delta l_{QM}=2(l/m)^{1/2}$ and $\delta l_{C}$ are
 the quantum mechanical and  the purely gravitational
parts, respectively (we set $c=\hbar=1$). The $\delta l_{C}$ was
computed from the Schwarzschild metric \cite{Myers-Perry}
surrounding the clock
\begin{equation}\label{SchwarzschildSolution(4+n)}
\begin{array}{l}
\fl ds^2  =  - (1 - C/r^{1 + n} )c^2 dt^2  + (1 + C/r^{1 + n} )
^{ - 1} dr^2  + r^2 d\Omega _{(2 + n)}^2~, \\
\fl C = r_{S(4 + n)}^{1 + n}  = 8(2 + n)^{ - 1} \pi ^{ - (1 + n)/2}
\Gamma [(3 + n)/2]G_{(4 + n)} m ~.\end{array}\end{equation} Here
$G_{(4+n)}$ is the gravitational constant in $(4+n)$-dimensional
spacetime. Using the optical path from some point $r_{0}>r_{S(4+n)}$
to a generic point $r$ (see, for example, \cite{Land-Lifsh})
\begin{equation}\label{OpticPathCase4}
\fl c\Delta t = \int_{r_0 }^r {\frac{{dr}}{{1 - \left( {{{r_{S\left(
{4 + n} \right)} } \mathord{\left/{\vphantom {{r_{S\left( {4 + n}
\right)} } r}} \right. \kern-\nulldelimiterspace} r}} \right)^{1 +
n} }}}  = \left( {r - r_0 } \right) + \int_{r_0 }^r
{\frac{{r_{S\left( {4 + n} \right)}^{1 + n} dr}}{{r^{1 + n}  -
r_{S\left( {4 + n} \right)}^{1 + n} }}}~,
\end{equation} the authors of \cite{Scar-Cas}
for the uncertainty due to curvature have found
\begin{equation}\label{ErrorDueToCurvature}
\fl \delta l_C  = \int_{a }^r {\frac{{r_{S\left( {4 + n} \right)}^{1
+ n} dr}}{{r^{1 + n}  - r_{S\left( {4 + n} \right)}^{1 + n} }}}
~.\end{equation}

From our point of view in order to take proper account of all
possible uncertainties in a distance measurement one must use the
following expression for the total uncertainty
\begin{equation}\label{}
\fl \delta l_{total}=2(l/m)^{1/2}+a+\int_{a}^r {\frac{{r_{S\left( {4
+ n} \right)}^{1 + n} dr}}{{r^{1 + n}  - r_{S\left( {4 + n}
\right)}^{1 + n} }}}~,
\end{equation} where additional term of the order of clock's size $a$ originates from the uncertainty in the registration of the returned
light signal in optical path measurement process. Obviously, in
order that the clock is not a black hole, one must suppose $a=\alpha
r_{S(4+n)}$, where dimensionless coefficient $\alpha > 1$.

Using the estimations of the integrals presented in \cite{Scar-Cas},
in the case $l > a \gg r_{S\left( {4 + n}\right)}$ for the total
uncertainty in distance measurement one gets \numparts
\begin{eqnarray}  \label{TotalUncertainty-a}\fl
\delta l_{tot}  = 2(l/m)^{1/2}+\alpha r_{S(4)}+
r_{S(4)} \ln (l/a)~~~~{\rm{in~ the~
case }}~~n = 0, \\\label{TotalUncertainty-b} \fl \delta l_{tot}  =
2(l/m)^{1/2}+{\alpha r_{S(4 + n)} + \left( {{{r_{S(4 + n)}^{1 + n} }
\mathord{\left/ {\vphantom {{r_{S(4 + n)}^{1 + n} } n}}
\right.\kern-\nulldelimiterspace} n}} \right)\left[ {a^{ - n}  - l^{
- n} } \right]~~~{\rm{in~ the~ case }}~~n > 1.}\end{eqnarray}
\endnumparts

In case $n=0$, i.e. $4$-dimensional spacetime, it is obvious from
(\ref{TotalUncertainty-a}) that the second and third terms in the
r.h.s. have similar dependences on the clock's mass, i.e. through
$r_{S(4)}$, and {\it a priori} we can not say which of this two
terms is of less order. The minimization of the total distance
uncertainty with respect to the clock's mass gives the result
$\delta l_{tot\_min}\sim [ll^{2}_{Pl(4)}]^{1/3}$ which complies with
the holographic principle ($l_{Pl(4)}$ is the Planck length in
$4$-dimensional spacetime). Now suppose that the second term in
(\ref{TotalUncertainty-a}) is missed, as it is in \cite{Scar-Cas}.
Even in this case the minimization still gives the total distance
uncertainty which complies with the holographic principle, and this
fact happens due to the similarity of this terms in dependence on
the clock's mass.

As regards the case $n \ge 1$ it is obvious from
(\ref{TotalUncertainty-b}) that the second and third terms in the
r.h.s. have different dependences on the clock's mass. But in the
taken case $l > a \gg r_{S\left( {4 + n}\right)}$ the uncertainty
due to curvature (the third term in the total uncertainty
(\ref{TotalUncertainty-b})) must be neglected as compared to the
second term associated with the clock's size, and thus subsequent
minimization of the total uncertainty with respect to clock's mass
gives the result $\delta l_{tot_min}\sim
[ll^{2+n}_{Pl(4+n)}]^{1/(3+n)}$ , which respects the holographic
principle ($l_{Pl(4+n)}$ is the Planck length in $(4+n)$-dimensional
spacetime). If in this case the second term in
(\ref{TotalUncertainty-b}) is missed (as it is in \cite{Scar-Cas}),
then the minimization gives an incorrect total uncertainty in
distance which does not comply with the holographic principle.

Thus our investigation states that the modified uncertainty
principle proposed in \cite{Scar-Cas} gives the uncertainty in
distance measurement which complies with the holographic principle
in spacetimes with extra dimensions. \ack{The author is indebted to
M. Gogberashvili, M. Maziashvili and B. Midodashvili for useful
conversations.}

\end{document}